\documentclass[oldversion]{aa} 
\usepackage{graphicx}
\usepackage{aalongtable}
\usepackage{txfonts}
\usepackage{rotating}
\usepackage{lscape}
\newcommand{\gsim}{\;\lower.6ex\hbox{$\sim$}\kern-7.75pt\raise.65ex\hbox{$>$}\;}
\newcommand{\lsim}{\;\lower.6ex\hbox{$\sim$}\kern-7.75pt\raise.65ex\hbox{$<$}\;}

\begin{document}
\title{Aluminium abundances in five discrete stellar populations of the globular
cluster NGC 2808\thanks{Based on observations collected at 
ESO telescopes under programme 094.D-0024}\fnmsep\thanks{
   Tables 1 and 3 are only available in electronic form at the CDS via anonymous
   ftp to {\tt cdsarc.u-strasbg.fr} (130.79.128.5) or via
   {\tt http://cdsweb.u-strasbg.fr/cgi-bin/qcat?J/A+A/???/???}}
 }

\author{
Eugenio Carretta\inst{1},
Angela Bragaglia\inst{1},
Sara Lucatello\inst{2}
Raffaele G. Gratton\inst{2},
Valentina D'Orazi\inst{2},
\and
Antonio Sollima\inst{1}
}

\authorrunning{E. Carretta et al.}
\titlerunning{Aluminium abundances in NGC~2808}

\offprints{E. Carretta, eugenio.carretta@oabo.inaf.it}

\institute{
INAF-Osservatorio di Astrofisica e Scienza dello Spazio di Bologna, Via Gobetti
 93/3, I-40129 Bologna, Italy
\and
INAF-Osservatorio Astronomico di Padova, Vicolo dell'Osservatorio 5, I-35122
 Padova, Italy}

\date{}

\abstract{We observed a sample of 90 red giant branch (RGB) stars in NGC~2808
using FLAMES/GIRAFFE and the high resolution grating with the set up HR21. These
stars have previous accurate atmospheric parameters and abundances of light elements. 
We derived aluminium abundances for them from the strong doublet Al {\sc i}
8772-8773~\AA\ as in previous works of our group. In addition, we were able to
estimate the relative CN abundances for 89 of the stars from the strength of a 
large number of CN features. When adding self consistent abundances from previous UVES spectra
analysed by our team, we gathered [Al/Fe] ratios for a total of 108 RGB stars in
NGC~2808. The full dataset of proton-capture elements is used to explore in
details the five spectroscopically detected discrete components in this globular
cluster. We found that different classes of polluters are required to
reproduce the (anti)-correlations among all proton-capture elements in the
populations P2, I1, and I2 with intermediate composition. This is in
agreement with the detection of lithium in lower RGB 
second generation stars, requiring at least two kind of polluters. To have
chemically homogeneous populations the best subdivision of our sample is into
six components, as derived from statistical cluster analysis. By comparing
different diagrams [element/Fe] vs [element/Fe] we show for the first time  that
a simple dilution model is not able to reproduce all the sub-populations in this
cluster. Polluters of different masses are required. NGC~2808 is confirmed to
be a tough challenge to any scenario for globular cluster formation.
}

\maketitle

\section{Introduction}

Proton-capture reactions involving the complete CNO, and the Ne-Na,
Mg-Al  cycles can simultaneously be at work in H-rich stellar interiors when
temperature is high enough (Denisenkov and Denisenkova 1989, Langer et al.
1993) in globular clusters (GCs), the first aggregates to appear in the forming
galaxy. 
These reactions produce Na from Ne and N from C and O when the temperature exceeds
about 30 MK. Conversion of Mg into Al occurs when H-burning occurs at temperature 
exceeding 65 MK. If material processed through these reactions is somehow given
back to the interstellar medium, we may observe significant deviations from the
typical low-Na, high-O plateau established in metal-poor halo stars (e.g.
Wheeler, Sneden, Truran 1989). These alterations are just those discovered by
the Lick-Texas group (see Kraft 1994 for a summary) in a good fraction of
GC stars (as reviewed by Gratton, Sneden, Carretta 2004).
They are only found among GC stars and in virtually all GCs  (e.g. Gratton et al. 2000; 
Carretta et al. 2009a,b; see references in Gratton, Carretta, Bragaglia 2012; 
Bragaglia et al. 2017, Bastian and Lardo 2017),
indicating that this nucleosynthetic signature is related to high density environments,
and it is not present e.g. among stars in dwarf spheroidal galaxies 
(see e.g. Carretta 2013, Lardo et al. 2016).

While these alterations are detected in currently evolving GC stars, $in$ $situ$ 
production is ruled out because temperatures in their interior is not high enough. 
In addition, the distribution of light element abundances is similar for giants
with deep convective envelopes and main sequence stars with negligible convective 
envelopes (and they are: e.g. Cannon et al. 1998, Gratton et al. 2001, Briley et 
al. 2002, 2004a,b, Harbeck et al. 2003, Cohen et al. 2002, 2005, D'Orazi et al. 
2010, Dobrovolskas et al. 2014). The nucleosynthetic site must then
be searched in stars of a previous stellar generation (Gratton et al.
2001) more massive than those currently evolving through the red giant branch (RGB)
of GCs.

Finally, these star-to-star abundance variations in light elements are typically $not$
accompanied by intrinsic variations in the iron content (constant in most GCs: 
e.g. Carretta et al. 2009c, but see e.g. Johnson et al. 2015 and Marino et al.
2015 for the growing class 
of iron complex GCs). The release of proton-capture elements likely did not occurred
during the epoch when supernova exploded in the clusters.

More than three decades of spectroscopic observations set the general background 
of the current paradigm for GCs. These old objects are not as simple as once thought. 
They formed in at least two close (time delay from a few to a few tens of Myr) 
bursts of star formation with second generation (SG) stars incorporating the 
proton-capture yields released at low velocity by some of the massive stars of 
the first generation (FG). Constraints from the chemical feedback available in 
FG stars, abundances of the fragile Li detected in SG stars together with matter processed 
at high temperature, and considerations from stellar nucleosynthesis call for a 
significant fraction of unprocessed, pristine gas mixed with ashes of H-burning to 
form all other stellar generations but the FG (Prantzos and Charbonnel 2006, D'Ercole,
D'Antona, Vesperini 2011).

While this general framework is broadly (though not universally, see e.g. Bastian and 
Lardo, 2015; Bastian, Cabrera-Ziri and Salaris 2015) accepted, many important features are still
hotly debated. For instance, it is not entirely clear which class of FG stars was
the main player to pollute the intra-cluster medium (e.g. massive binaries, de
Mink et al. 2009; massive, fast rotating single stars, FRMS, Decressin et al.
2007, intermediate mass AGB and super AGB stars, Ventura et al. 2001, D'Ercole
et al. 2010, 2012) or whether more than one class was contributing to the SG. 
The ensemble of nucleosynthesis constraints is not completely satisfied by
any of the proposed candidate polluters (Prantzos, Charbonnel, and Iliadis
2017).

The ubiquitous Na-O anti-correlation alone, so widespread in GCs to be suggested as the
main chemical feature characterizing the essence of a genuine Galactic 
GC (Carretta et al. 2010a; see Villanova et al. 2013 for the notable exception
of Rup 106), is not sufficient to provide strong constraints: several classes
of candidate polluters were able to develop temperature high enough to manage
the proton-capture reactions depleting O and enhancing Na.

More insight is obtained by considering also cycles requiring much higher
temperatures, such as the Mg-Al chain. In general, these reactions require
higher masses for the polluters. While indications for the activation of this
cycle in the GC polluters date back to Kraft et al. (1997), most early studies
only  considered a small number of clusters (e.g. Johnson et al. 2005) due to
limitations in single slit spectrographs. Once efficient multi-object
spectrographs became available, much  more extensive survey became possible.
Carretta et al. (2009b) performed an homogeneous scrutiny of high resolution
UVES-FLAMES spectra of a limited number  of RGB stars (max 14) in 17 GCs. They
showed that an efficient production of Al  only occurred in massive and/or
metal-poor GCs. This finding was later confirmed by the near-infrared 
APOGEE survey in 10  northern GCs (M\'esz\'aros et al. 2015).
This indicates that the involved  high temperatures were not
reached in the H-burning regions of the polluters for  the smallest/metal-rich
GCs, as instead it happened for the CNO and Ne-Na cycles.  This indicated that the
typical mass of the polluters correlates with the mass of the cluster. In
addition, since the Al content is one order of magnitude less  than Mg in the
Sun and in metal-poor stars, this means that when a sizeable fraction  of the
original Mg is transformed into Al, its abundance may change by a huge factor. 
This offers an extraordinary resolution to probe processes occurring during the 
formation of GCs allowing e.g. much better separation of stars into different 
chemical homogeneous groups.

We begun to systematically exploit this approach by observing large samples of
RGB stars (about 100) in NGC~6752 (Carretta et al. 2012a),
NGC~1851 (Carretta et al. 2012b), 47 Tuc and M~4 (NGC~6121: Carretta et al.
2012c) with FLAMES/GIRAFFE (Pasquini et al. 2002) and the efficient set up HR21,
whose spectral range includes the strong Al {\sc i} 8772-8773~\AA\ doublet.
Briefly, leaving aside the iron complex cluster NGC~1851, we studied the
complete and large dataset of homogeneous O, Na, Mg, Al abundances with a $k-$means
algorithm (Steinhaus 1956, MacQueen 1967) of cluster analysis, finding three
discrete populations in both NGC~6752 and 47~Tuc. These two GCs however differ
because in the latter it is possible to reproduce the group with intermediate
(I) abundances by mixing the compositions of the primordial (P) and extreme (E)
populations, whereas for NGC~6752 a simple dilution model cannot account
simultaneously for all the P, I, E components. This implies the action of at least
two different classes of polluters. The existence of discrete stellar components
in NGC~6752 and 47~Tuc is also supported by photometric studies, based on larger
samples of stars (Carretta et al. 2011, Milone et al. 2013; and Milone et al.
2012, respectively).

The discovery and chemical characterization of possible distinct groups in the
multiple stellar populations of a GC is crucial to reconstruct the
possible sequence of early star formation in these objects: whether polluters
of the same class but different mass ranges contributed, as suggested by the
three groups following the dilution track in 47 Tuc, or rather two
different kind of contributors, acting at different times, as likely in
NGC~6752.

In the present paper we extend our Al studies to the peculiar GC NGC~2808, where
five discrete components with distinct chemical compositions were discovered 
in Carretta (2015), supporting the analogous finding from photometry (Milone et al.
2015, who discussed the existence of up to seven populations). By more than tripling the sample of giants with Al determinations in this
cluster (Carretta 2014) we gather a set of proton-capture elements (O, Na,
Mg, Al, Si, with also estimates of N content) and apply again the cluster
analysis to search for chemically homogeneous sub-populations and provide more
stringent constraints on the formation scenario for this cluster 
(see D'Antona et al. 2016).

The paper is organized as follow: in Sect. 2 will be presented the observational
data, analyzed to derive abundances in Sect. 3, and discussed in Sect. 4. In
Sect. 5 we show results of the statistical cluster analysis, and in Sect. 6 we
present final discussion and conclusions.

\section{Observations and ancillary data}

Our sample of first ascent red giant stars in NGC~2808 was observed in
Service mode with  FLAMES@VLT-UT2 on 2015, January 21, with an exposure time of
1 hour at airmass 1.314 with the high resolution GIRAFFE setup HR21 (R=17,300
and spectral interval from about 8484~\AA\ to about 9001~\AA).
As in previous works on Al in NGC~6752 (Carretta et al. 2012a), M~4 and 47~Tuc
(Carretta et al. 2012c) we adopted the same configurations for fiber positioning
that was used to observe stars in NGC~2808 (Carretta et al. 2006) with HR11.
This choice permits to maximize the number of giants with measured
abundances of Na, because the five distinct populations found by Carretta (2015) 
were best detected using a combination of Na and Mg.

We obtained spectra for 90 members. Data reduced by
ESO personnel with the dedicated pipeline (spectra de-biased, flat-fielded,
extracted and wavelength calibrated) were retrieved, sky subtracted and 
shifted to zero radial velocity using IRAF\footnote{IRAF is the Image Reduction
and Analysis Facility, a general purpose software system for the reduction and
analysis of astronomical data. IRAF is written and supported by the IRAF
programming group at the National Optical Astronomy Observatories (NOAO) in
Tucson, Arizona. NOAO is operated by the Association of Universities for
Research in Astronomy (AURA), Inc. under cooperative agreement with the National
Science Foundation.}. 

\setcounter{table}{0}
\begin{table*}
\centering
\caption[]{Information on stars observed in NGC~2808.  The complete
Table is available electronically only at CDS.}
\begin{tabular}{rccrrc}
\hline
star      & RA  & DEC & S/N & RV Hel.     & V       \\
          &     &     &     & kms$^{-1}$ & mag     \\
\hline        
\hline
  7183  &  138.011958 & -64.826111 & 157.8 &  	104.47 &  14.854  \\
  7315  &  137.994083 & -64.824944 & 162.9 & 	 99.44 &  14.683  \\
  7788  &  137.988291 & -64.820694 &  99.8 & 	 97.64 &  14.870  \\
  8198  &  137.954041 & -64.816527 & 115.5 & 	 98.10 &  15.340  \\
  8204  &  137.994041 & -64.816444 & 130.0 & 	110.22 &  15.119  \\
  8603  &  138.058541 & -64.811916 & 195.3 & 	110.15 &  14.432  \\
  8739  &  137.963333 & -64.810416 & 196.9 & 	109.36 &  14.288  \\
  8826  &  138.040833 & -64.809222 & 115.0 & 	 96.85 &  15.033  \\
\hline
\end{tabular}
\label{t:info28}
\end{table*}

\begin{figure}
\centering
\includegraphics[scale=0.40]{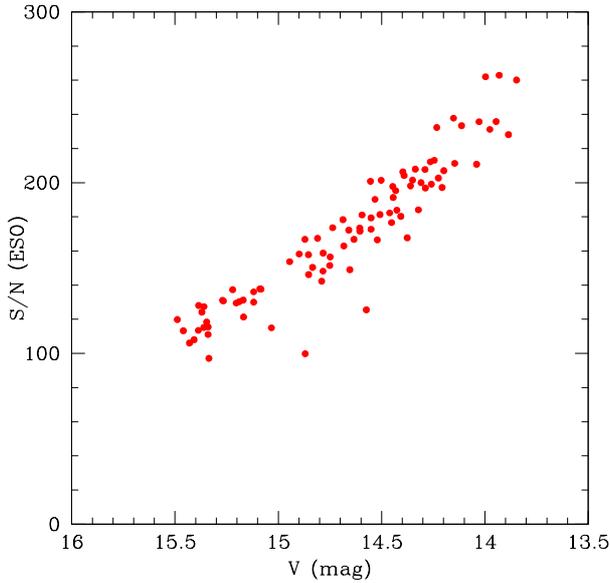}
\caption{S/N ratios from the ESO FLAMES-GIRAFFE pipeline as a function of the
$V$ magnitudes for our sample in NGC~2808}
\label{f:sn28vsV}
\end{figure}

Our target stars are red giants in the magnitude range $V=13.8-15.5$. Star
identification, coordinates and Johnson $V$ magnitudes (Carretta et al. 2006)
are reported in Table~\ref{t:info28}, together with the heliocentric radial
velocity (RV) we derived. Also listed in this Table is the S/N ratio provided by
the ESO pipeline as the mean value across the spectrum. The S/N basically scales
as a function of the magnitude, apart from a couple of outliers 
(Fig.~\ref{f:sn28vsV}). The median S/N is 172 for our sample.

Atmospheric parameters (effective temperature, surface gravity, model
metallicity and microturbulent velocity) for all stars were adopted from
Carretta (2015), since in that work they were updated, with respect to the values
in Carretta et al. (2006), to the homogeneous system used in all
the other GCs in our FLAMES survey for the Na-O anti-correlation. The abundance
ratios [O/Fe], [Na/Fe], [Mg/Fe], and [Si/Fe]\footnote{We adopt the usual spectroscopic notation, $i.e.$  for any
given species X, [X]=  $\log{\epsilon(X)_{\rm star}} - \log{\epsilon(X)_\odot}$
and   $\log{\epsilon(X)}=\log{(N_{\rm X}/N_{\rm H})}+12.0$\ for absolute number 
density abundances.} that, together with Al, complete the
dataset of proton-capture elements for NGC~2808, were also taken from Carretta
(2015).

\section{Analysis and derived abundances in NGC~2808}
 
Our methods to derive abundances of Al closely followed the procedure described
in Carretta et al. (2012a,b,c).

\begin{figure}
\centering
\includegraphics[scale=0.40]{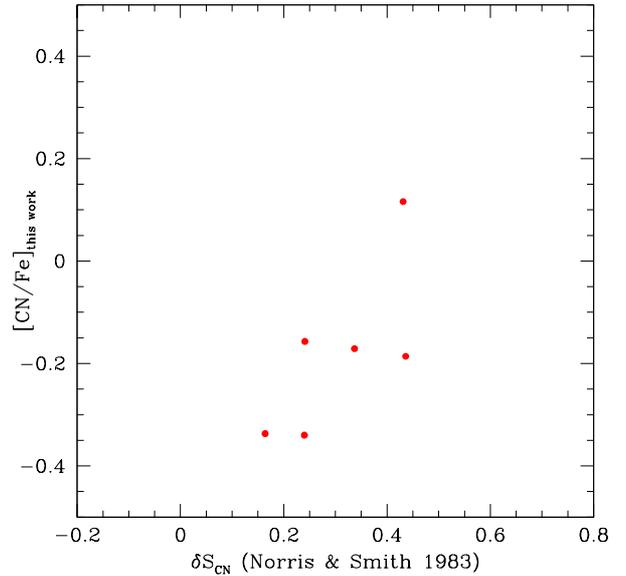}
\caption{Comparison of our estimates of [CN/Fe] in NGC~2808 with the cyanogen
excesses derived in Norris \& Smith (1983) for 6 stars in common.}
\label{f:dscn}
\end{figure}

\subsection{Estimates of nitrogen abundances}

The ubiquitous CN lines present in the spectral range of HR21 may give a certain
degree of contamination, spuriously affecting abundances derived from the Al
features of interest. At the same time, they offer the opportunity to provide
rough estimates of the abundances of N, another light element involved in the
proton-capture reactions in H-burning.

We summed the spectra of the 18 coolest stars in our sample 
(T$_{\rm eff} < 4300$ K) to obtain a master spectrum of very high S/N. On this
co-added spectrum we determined 8 regions to be used as local reference continuum
and 20 other regions dominated by CN features. The spectral ranges for the
continuum regions coincide with those used in Carretta et al. (2012c), whereas
the selected features for CN slightly differ from those adopted for 47~Tuc, due
to the different metallicity and quality of spectra. All the regions used for
NGC~2808 are listed in Table~\ref{t:feature}.

\setcounter{table}{1}
\begin{table*}
\centering
\caption[]{List of regions used to estimate local continuum and fluxes of CN
bands}
\begin{tabular}{lccclccc}
\hline
      & $\lambda_{in}$  & $\lambda_{fin}$  & $\Delta \lambda$ &  & $\lambda_{in}$  & $\lambda_{fin}$  & $\Delta \lambda$ \\
      &    \AA          &      \AA         &     \AA          &  &	 \AA	   &	 \AA	      &      \AA	 \\
\hline        
  CN  & 8704.86 & 8705.63 & 0.77 &    CN  & 8813.42 & 8814.01 & 0.59 \\
  CN  & 8706.28 & 8707.50 & 1.22 &    CN  & 8815.63 & 8816.52 & 0.89 \\
  CN  & 8718.43 & 8719.20 & 0.77 &    CN  & 8819.04 & 8820.32 & 1.28 \\
  CN  & 8720.13 & 8721.63 & 1.50 &    CN  & 8822.12 & 8823.38 & 1.26 \\
  CN  & 8722.32 & 8723.32 & 1.00 &    CN  & 8830.32 & 8831.44 & 1.12 \\
  CN  &	8759.22 & 8760.22 & 1.00 &    CN  & 8835.07 & 8836.86 & 1.79 \\ 
  CN  & 8767.24 & 8768.24 & 1.00 &  cont  & 8739.29 & 8739.71 & 0.42 \\ 
  CN  & 8769.96 & 8771.34 & 1.38 &  cont  & 8746.22 & 8746.53 & 0.31 \\ 
  CN  & 8774.84 & 8776.12 & 1.28 &  cont  & 8765.41 & 8765.60 & 0.19 \\ 
  CN  & 8782.88 & 8783.82 & 0.94 &  cont  & 8771.32 & 8771.65 & 0.33 \\ 
  CN  & 8788.92 & 8789.61 & 0.69 &  cont  & 8791.26 & 8791.46 & 0.20 \\ 
  CN  & 8794.02 & 8794.76 & 0.74 &  cont  & 8791.74 & 8792.12 & 0.38 \\ 
  CN  & 8801.77 & 8802.65 & 0.88 &  cont  & 8794.76 & 8795.32 & 0.56 \\ 
  CN  & 8812.42 & 8813.42 & 1.00 &  cont  & 8834.69 & 8835.14 & 0.45 \\ 
 
\hline
\end{tabular}
\label{t:feature}
\end{table*}

For each star, three synthetic spectra were computed with the package ROSA
(Gratton 1988) using [N/Fe]=-0.5, +0.25 and +1.0 dex, the Kurucz (1993) grid of
model atmospheres (with the overshooting option on), atmospheric parameters from
Carretta (2015), the same line list as in Carretta et al. (2012b), and assuming
[C/Fe]=0. This is an arbitrary assumption, due to our ignorance of the
actual C pattern in giants of NGC~2808. As a consequence, we are actually
measuring the abundance of the C$\times$N product. Were the actual C content
different from the assumed value, the N abundance would be different as to keep
constant the sum [C/Fe] + [N/Fe].

A weighted reference continuum was derived from all the  8 regions using weights
equal to the width of each region. For each CN feature the average flux within
the in-line region was measured, and the associated CN abundance was obtained by
comparing the normalized flux with the fluxes measured in the same way on the
three synthetic spectra. Finally, we applied a 2.5 $\sigma$-clipping
to the average abundance from individual features in each star,  after
discarding features that visual inspection revealed affected by spikes.

\setcounter{table}{2}
\begin{table*}
\centering
\caption[]{Group classification and derived abundances for RGB stars in 
NGC~2808. The complete Table is available electronically only at CDS.}
\begin{tabular}{rccrcrcr}
\hline
   star & S$^1$ & group$^2$ & [Al/Fe] &    err$^3$  &    [CN/Fe]  &  rms    &  Nr$^4$     \\
\hline
\hline
   7183 & G &	 2  &    0.091&  0.020  &    -0.219  &  0.111  & 19   \\
   7315 & G &	 1  &   -0.236&  0.021  &    -0.241  &  0.105  & 20   \\
   7536 & U &	 1  &    0.201&  0.007  &            &  9.999  &  0   \\
   7558 & G &	 1  &         &         &            &         &  0   \\
   7788 & G &	 2  &   -0.169&  0.033  &    -0.157  &  0.103  & 18   \\
   8198 & G &	 1  &   -0.369&  0.030  &    -0.171  &  0.172  & 18   \\
   8204 & G &	 1  &   -0.131&  0.024  &    -0.294  &  0.115  & 17   \\
   8603 & U &	 1  &    0.030&  0.024  &    -0.340  &  0.084  & 17   \\
\hline
\end{tabular}
\label{t:tab28}
\begin{list}{}{}
\item[1-] G = GIRAFFE; U = UVES
\item[2-] group classification from Carretta (2015)
\item[3-] for abundances from UVES this is rms scatter of the mean from the two
lines at 6696-98~\AA.
\item[4-] number of used CN features.
\end{list}
\end{table*}

We obtained abundances for 89 stars using on average 18 features, finding 
on average [CN/Fe]$=-0.191 \pm 0.014$ dex (rms=0.129 dex) in NGC~2808.
Individual values of CN abundances are listed in Table~\ref{t:tab28}.

In Fig.~\ref{f:dscn} we compare our estimates of the [CN/Fe] ratios to the
cyanogen excesses derived by Norris and Smith (1983) for 6 giants of NGC~2808 in 
common between their study and the present work. The good correlation supports
the conclusion that our procedure provides a good estimate of the CN content, 
despite the arbitrary assumption about the unknown abundance of C. As a
final $caveat$, we recall that the total CN range may be underestimated since it
is well documented that C abundances are higher (by about 0.3 dex) in CN-weak,
SG stars. However, in M~4, a GC with similar metallicity, the C abundances for
SG, Na-rich stars and FG, Na-poor stars only differ by 0.06 dex, on average (see
discussion and references in Carretta et al. 2012c). In the following, to avoid
any misunderstanding, we will indicate with [CN/Fe] our derived abundance
ratios.

\begin{figure*}
\centering
\includegraphics[scale=0.45]{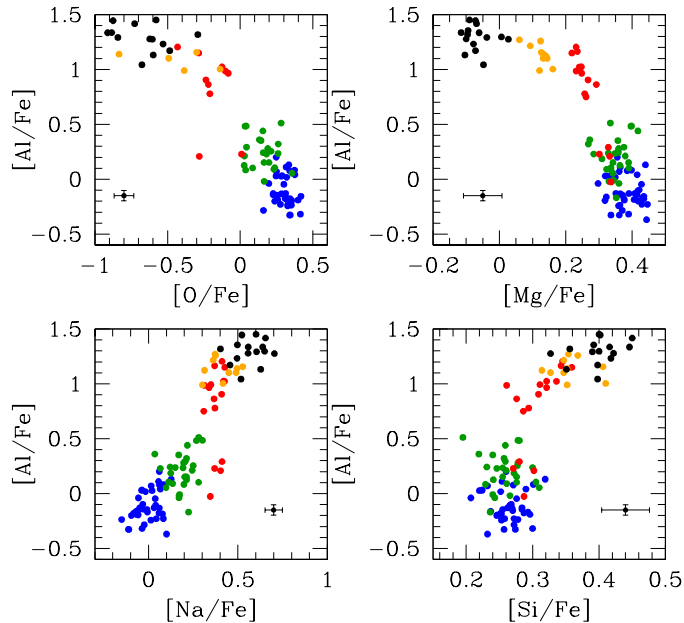}
\caption{Abundance ratios [Al/Fe] as a function of the proton-capture 
elements O, Na, Mg, Si in RGB stars of NGC~2808. Different
colours indicate the five populations as defined in Carretta (2015) using the
Mg-Na plane (population P1: blue, P2: green, I1: red, I2: orange, E: black). In 
each panel the star-to-star error bars are indicated.}
\label{f:light3t28def}
\end{figure*}

\begin{figure*}
\centering
\includegraphics[scale=0.45]{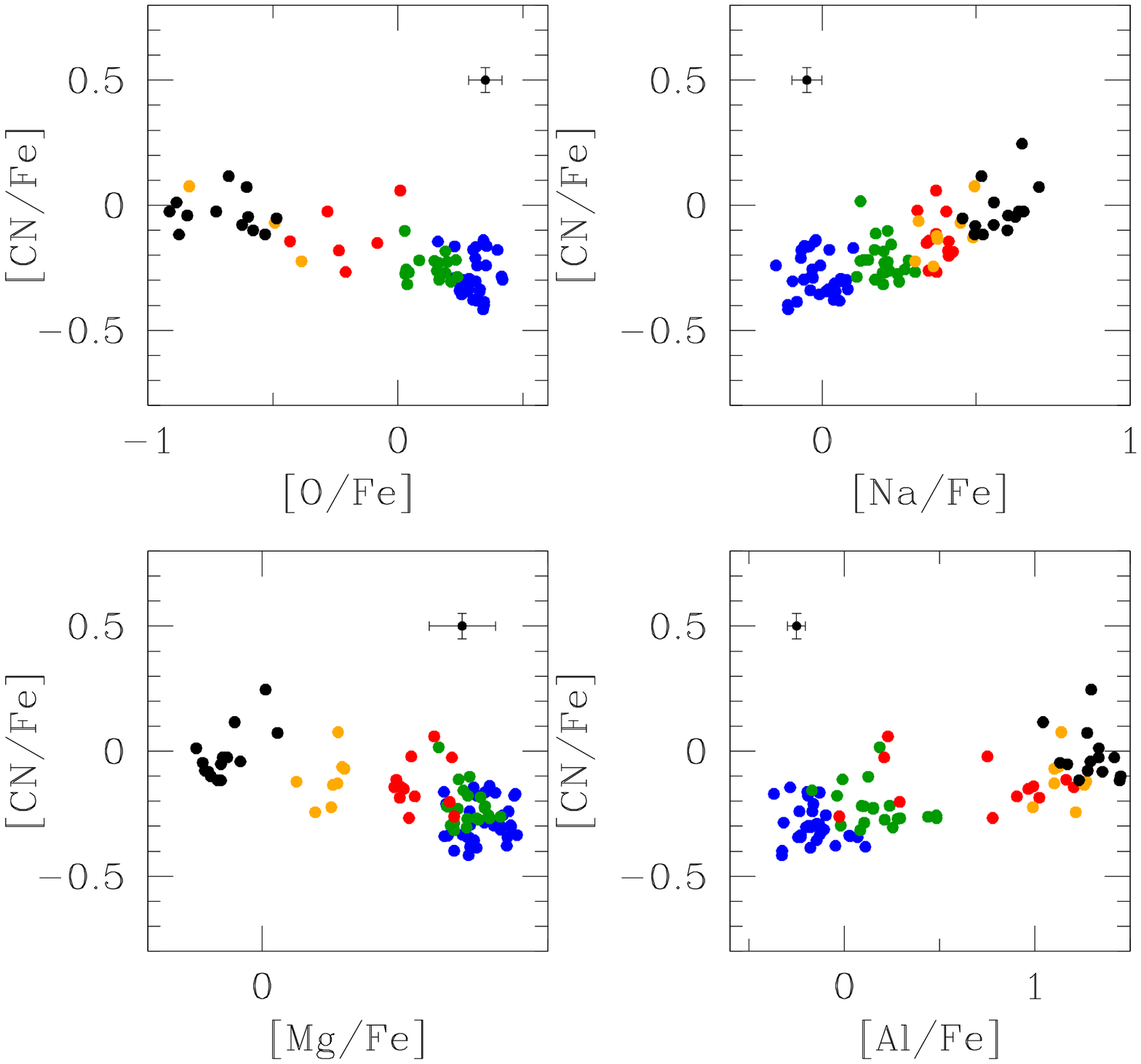}
\caption{As in Fig.~\ref{f:light3t28def} for the derived [CN/Fe] abundance
ratios.}
\label{f:light4t28def}
\end{figure*}

\subsection{Aluminium abundances}

After estimating the amount of the contaminant CN, we computed for each star
three synthetic spectra with [Al/Fe]=-0.5, 0.5, and 1.5 dex to derive Al
abundances measuring fluxes in the in-line region from  8772.25 to 8774.57~\AA\,
and using the regions 8769.09-8770.00~\AA\ and 8776.40-8777.70~\AA\ for the
local reference continuum, again individuated on the co-added, high S/N master
spectrum. The contribution from these two regions was averaged with weights
given by the number of pixels (i.e. the widths of these intervals).

We first inspected the Al lines and the selected continuum regions on plots
of the spectra, to be sure that none were affected by a defect, otherwise the
region was disregarded. We deemed not possible to derive Al
abundances for four stars. Then we interpolated the observed normalized fluxes
among those from the above synthetic spectra, obtaining the Al abundance for
each star.

Our aim is to gather the largest possible sample of RGB stars in NGC~2808 with
homogeneous abundances of light elements. To safely add the sample with high
resolution UVES spectra re-analyzed in Carretta (2014,2015), who used the Al
doublet at 6696-98~\AA, we checked possible offsets in Al using nine stars with both
doublets available. From these stars we found that on average the
difference is [Al/Fe](669-977 nm) = $+0.068 \pm 0.076$ dex, with $rms=0.229$,
compatible with no significant offset.

As in our previous papers the Al abundances are in LTE and our adopted solar
abundance for Al is 6.23. In the following we then
adopt whenever possible [Al/Fe] abundances derived from UVES spectra by Carretta
(2014, 2015). Final Al abundances for giants in  NGC~2808 are listed in
Table~\ref{t:tab28}. We were able to derive an Al  value for 86 stars observed
with FLAMES/GIRAFFE and HR21. Adding the 31 stars with determinations from UVES
(Carretta 2014) and taking into account nine stars in common we ended up with
108 RGB stars with homogeneous Al abundances.

Recently, D'Orazi et al. (2015) derived Al abundances for a sample of 65 stars
in NGC~2808 using the near infrared doublet at 6696-98~\AA. A direct,
star-to-star comparison is not possible, since that work was mainly aimed to
derive Li abundances, so their sample is made of lower RGB stars fainter
than any in the present sample. The average value they found ([Al/Fe]=+0.32
$\sigma=0.39$ dex) compares well with the mean [Al/Fe]=+0.46 $\sigma=0.47$ dex
for the sample of 31 giants with UVES spectra (Carretta 2014) and the average
value [Al/Fe]=+0.41 $\sigma=0.63$ dex from 86 stars with FLAMES HR21 spectra.
Differences may be explained by small differences in the derivation of
atmospheric parameters, e.g. the final temperatures are derived here and in
Carretta (2014) as a function of the 2MASS $K$ magnitude, whereas D'Orazi et al.
used the Johnson $V$ magnitude. They found that stars in NGC~2808 are split
into three main components, according to the Al abundances.

UVES spectra of seven asymptotic giant branch (AGB) stars in NGC 2808 were
analyzed by Marino et al. (2017). Once Al and Fe are corrected to our scale 
for offsets due to the adopted  solar abundances, they derived an average ratio
[Al/Fe]=+0.63 $\sigma=0.40$ dex. Note that this last sample does not
include the most Al-rich population, since apparently these stars, being
probably also He-enriched, do miss the AGB phase.

Part of the difference may be due to (neglecting) corrections from
departures from the LTE assumption. Nordlander and Lind (2017) claim that on the
lower RGB in metal-poor GCs abundance corrections for Al from optical and
near-IR lines depends on the Al abundance, leading to a possible compression of
the abundance scale. When the analysis is restricted to a more limited
evolutionary phase, as the upper RGB in the present work, this effect should not
be of concern. We interpolated the corrections for 1D models of their benchmark
stars in their Table 1 and estimated that the NLTE Al abundance for a star with
the average parameters in our sample (4442/1.42/-1.13/1.57) is only 0.06 dex
larger than the LTE value. NLTE corrections would be -0.09 and +0.06 dex for
our coolest and warmest stars. These corrections are derived from a
large number of Al lines, from UV to IR. However, from figures 13 and 14 in
Nordlander and Lind (2017) it is possible to conclude that NLTE corrections for
the abundances obtained only from the two lines used here are always less than
about 0.1 dex. The same approximatively holds for the analysis by D'Orazi et al.
and Marino et al. We preferred not to apply any correction for NLTE, to be fully
consistent with Al analysis in the other paper of this series. The impact on
star-to-star comparison in NGC~2808 is negligible.

\subsection{Errors}

To estimate star-to-star errors due to uncertainties in the adopted atmospheric
parameters we followed again closely the approach used in Carretta et al.
(2012c). We need to estimate the internal errors in atmospheric parameters and
evaluate the sensitivity of abundances to the adopted parameters. Internal
errors associated to the atmospheric parameters are simply taken from Carretta
(2015). They were estimated as 5 K, 0.041 dex, 0.026 dex, and 0.08 kms$^{-1}$
for T$_{\rm eff}$, $\log g$, [A/H], and $V_t$, respectively (second line in
Table~\ref{t:sensitivity}). To evaluate the sensitivity of Al to changes in
atmospheric parameters we considered star 7315, whose effective temperature
(4459 K) is very close to the mean T$_{\rm eff} = 4456$ K of the sample. The
analysis was then repeated by changing each time a single parameter by amounts
given in the first line of Table~\ref{t:sensitivity}, whereas the other were
fixed.

\setcounter{table}{3}
\begin{table*}
\centering
\caption[]{Sensitivities of Al to variations in the atmospheric
parameters and to errors in fluxes, and errors in 
abundances [A/Fe] for stars in NGC~2808}
\begin{tabular}{lrrrrrrr}
\hline
Element     & [CN/Fe]  & T$_{\rm eff}$ & $\log g$ & [A/H]   & $v_t$    & flux     & Total     \\
            &         &      (K)      &  (dex)   & (dex)   &kms$^{-1}$& (dex)   &Internal    \\
\hline        
Variation&   +0.200    &  50           &   0.200   &  0.100   &  0.10    &         &         \\
Internal &   +0.018    &   5           &   0.041   &  0.026   &  0.08    & 0.019   &         \\
\hline
$[$Al/Fe$]${\sc i}&  0.062 & $-$0.025      &$-$0.007   & 0.094 &	+0.003 &   &0.036    \\
\hline

\end{tabular}
\label{t:sensitivity}
\end{table*}

A second source of internal errors comes from errors in flux measurements. To
evaluate this component we estimated photometric errors from the S/N ratio of
the spectra and from the width within each of the reference continuum and the 
in-line regions (and then the number of pixels used, see Carretta et al. 2012c). We
then computed Al abundances with the adopted procedure, but using the new value
of the Al line strength index that is the sum of the original value and its
error. The comparison with original values provides the errors in [Al/Fe] due to
flux measurement errors,  listed in Table~\ref{t:tab28} for individual stars.
On average, the impact of this source of error is $+0.019 \pm 0.001$ dex
($rms=0.005$ dex, from 86 stars).

The same approach was applied to one of the CN features used to 
estimate the CN abundance. The average error is +0.018 dex, a conservative
estimate, since final CN abundances were obtained using up to 18 features, on
average. The corresponding impact on Al abundances is 0.006 dex.

The star-to-star (internal) errors in the Al abundances are obtained by summing
in quadrature all the contributions due to atmospheric parameters and flux
measurements (both in Al and CN features): the typical internal error is 0.036
dex.

\section{Proton-capture elements in NGC~2808}

The pattern of abundances of proton-capture elements among red giants in
NGC~2808 is summarized in Fig.~\ref{f:light3t28def} and 
Fig.~\ref{f:light4t28def}, where we plotted the abundance ratios derived here
([Al/Fe] and [CN/Fe]) as a function of proton-capture elements. The colour-coding
follows the five groups defined by Carretta (2015) using Mg and Na abundances. They
are the populations with primordial composition (P1, blue; P2, green), two
groups with intermediate composition (I1, red; I2, orange), and finally a
component with extremely modified composition (E, black).

From these figures we see that the newly derived abundances nicely follow the
overall pattern of the (anti-)correlations generated by the network of
proton-capture reactions. Al is anti-correlated with species depleted in
H-burning, like O and Mg, and is correlated with Na, Si, and CN, whose abundances
are enhanced in this nuclear processing. The same holds also for CN.

Four stars classified in the I1 group according to their Mg, Na abundances have
[Al/Fe] values too low, that clearly shift them to the P2 component. After
discarding one of the continuum region in one star and the Al line at 8772~\AA\
in another (due to spikes on the feature), there is no evident reason to doubt
of the derived Al abundances. The position of these four stars in the Na-Mg
plane (see Carretta 2015) is somewhat intermediate between the P2 and I1 group
and it is possible that they should be associated to the former, although their
Na values are on average as large as those of the other stars in the I1 component. 

\begin{figure}
\centering
\includegraphics[scale=0.40]{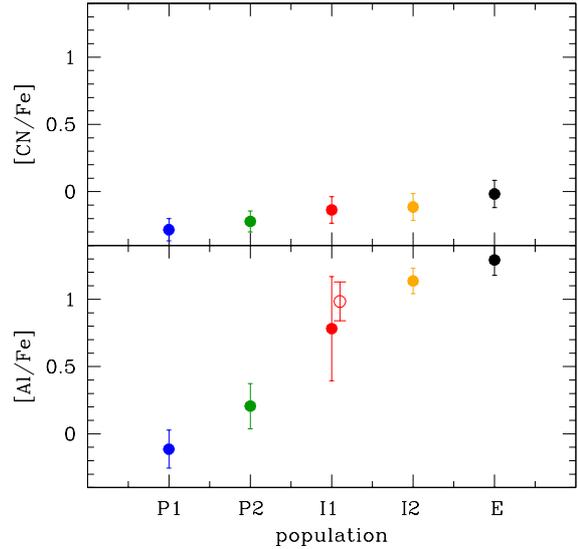}
\caption{Average values of the [CN/Fe] and [Al/Fe] abundance ratios in the five
discrete groups in NGC~2808. The error bar is the rms scatter of the mean. The
empty red circle represents the mean value for [Al/Fe] obtained by excluding the
four stars with too low Al values (see text).}
\label{f:trend2}
\end{figure}

\begin{figure}
\centering
\includegraphics[scale=0.40]{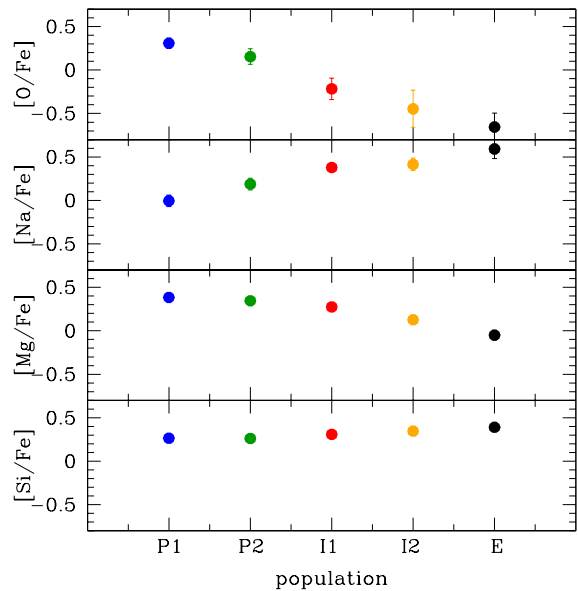}
\caption{As in Fig.~\ref{f:trend2} for O, Na, Mg, Si from Carretta (2015). In
many cases the rms scatter is within the point symbol.}
\label{f:trend4}
\end{figure}

Even taking into account these possible outliers, there is a clear progression
in the content of light elements among the sub-populations of stars detected in
NGC~2808, as shown in Fig.~\ref{f:trend2}, where we plotted the average values
for the elements analyzed in the present work, CN and Al. The mean values of
species enhanced in proton-capture reactions increase going from the P1, P2
groups up to the highest values in the component E. The peak-to-valley difference
in CN between P1 and E is not dramatically large (about 0.27 dex, although
this range may be somewhat underestimated, see Sect.3.1), but the
average level of Al increases by a full 1.4 dex between the two populations.
Allowing for the possibility that the four stars above are spuriously attributed
to the I1 group, the trend is still present, simply the rms scatter is reduced.
The trends for CN and Al complete those observed in Carretta (2015) for O, Na,
Mg, and Si. In Fig.~\ref{f:trend4} we plot the average values in the five
populations of NGC~2808.

As in Carretta (2015) we used the Student's and Welch's tests to check if the
average values for the five groups are significantly different from each other.
The average values for CN and Al are listed in Table~\ref{t:medieNAl} where 
we also report the mean values of O, Na, Mg, and Si from
Carretta (2015), for convenience of the readers.
For each group combination, the two-tail probability value for the mean [Al/Fe]
ratio never exceed $1.1 \times 10^{-3}$, usually being much smaller or even
zero. For CN, the probability values are generally low, indicating that mean
values are actually different, and not by mere random chance. The only exception 
is represented by the I1-I2 combination, with a two-tail $p=0.63$. This is
graphically supported by Fig.~\ref{f:trend2}, where the trend of the average CN
seems to show a flattening, and by Fig.~\ref{f:light4t28def}, where the two
populations seem to be intermingled (orange and red points).

\begin{figure}
\centering
\includegraphics[scale=0.40]{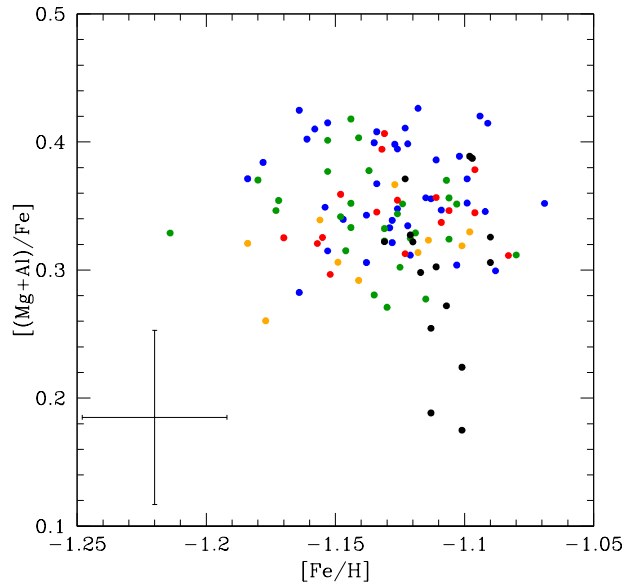}
\caption{Sum Al+Mg as a function of metallicity in the RGB stars in NGC~2808.
The colour coding is as in previous figures for stars of different populations.
Average internal errors are shown.}
\label{f:summgal}
\end{figure}

Anyway, it is clear that in NGC~2808 we see five discrete populations of RGB
stars showing different levels of Al, increasing when other elements enhanced in
proton-capture reactions simultaneously increase (and the corresponding depleted
species decrease). To test if this is a consequence of nucleosynthesis we simply
checked if the sum of Al+Mg remains constant, as it should be if all the Al were
produced by conversion of Mg in the Mg-Al cycle during H-burning at high
temperature. The result is illustrated in Fig.~\ref{f:summgal}, where the sum
[(Al+Mg)/Fe] is plotted as a function of the metallicity, adopting the above
colour coding for stars in each sub-population.
We found that the sum is constant: on average, $+0.342\pm 0.004$ dex, with
$\sigma=0.047$ dex (108 stars), to be compared with a total internal error of
0.068 dex in this sum. The production of Al is matched to a corresponding
consumption of Mg, and this result is only negligibly affected by the three
outliers visible at low values in Fig.~\ref{f:summgal}. These three objects
cannot be explained by leakage from the Mg-Al cycle on $^{28}$Si (Karakas and
Lattanzio 2003), since their Si abundance is high, but not the highest in our
sample. The same holds for their [K/Fe] values (Mucciarelli et al. 2015),
excluding that their position in Fig.~\ref{f:summgal} is due to neglecting
effects of H-processing at very high temperatures.

Next, we examined the case where a single class of polluters converting Mg into
Al may be able to reproduce the abundance pattern of the five groups through a
simple dilution model (see e.g. Carretta et al. 2009,a,b). In this case, the
abundances of individual stars are obtained by diluting the pure nuclearly 
processed matter of putative polluters with variable (increasing) fractions of
unpolluted gas having primordial composition.

\begin{figure}
\centering
\includegraphics[bb=17 150 344 716, clip, scale=0.59]{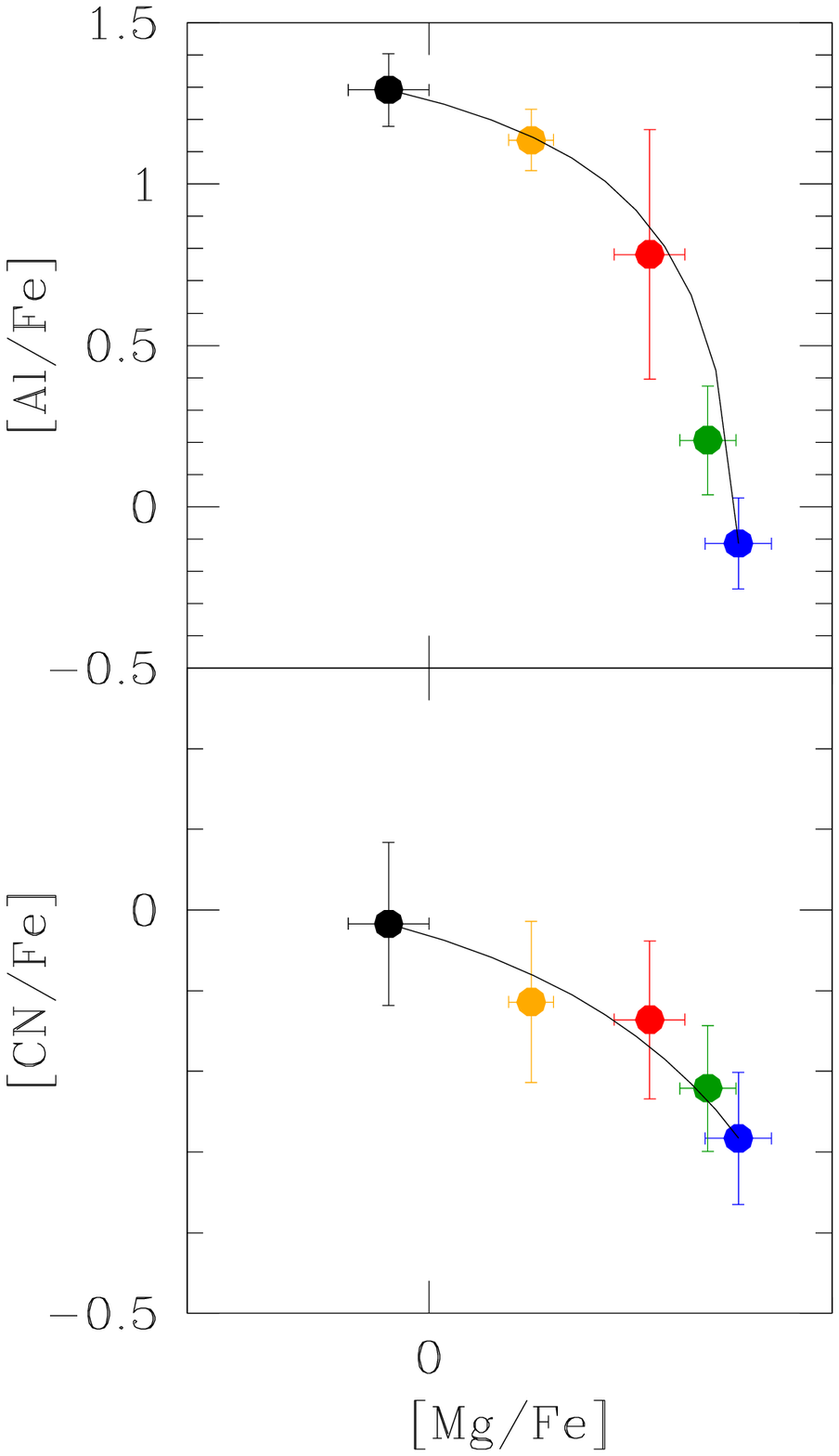}
\caption{Upper panel: average abundances of Al (this work) and Mg (from Carretta
2015) for the five sub-populations in NGC~2808. A simple dilution model is
superimposed to the data. Error bars are rms scatter. Lower panel: the same for
CN (from the present work).}
\label{f:a18}
\end{figure}

Results are shown in Fig.~\ref{f:a18} for the elements whose abundances were
derived in the present study: Al (upper panel) and CN (lower panel). 
Apparently, in both
cases a simple dilution model seems to be a satisfactory match for all the five
groups, within the observed scatter. Note that the lower panel is actually a
link between the O-N and the Mg-Al cycles, occurring a different temperatures,
yet the CN-Mg anti-correlation is rather well reproduced.

However, we caution that if the average abundances of the five
populations lie along a dilution model, this is a necessary, but not sufficient
condition. These plot are with logarithmic quantities, whereas those actually
variyng are linear quantities (the number of atoms of each species), and this
may give the impression of a good fit along a dilution curve. To verify this
issue, we adopted the approach used in Carretta et al. (2012a). If a
population/group with intermediate composition is obtained by diluting the
material with extreme E composition with pristine gas (of P composition), then
we should have: 
$$A(el)_I = A(el)_P + d \times [A(el)_E -A(el)_P] $$
where $d$ is the fraction of the material with E-like composition in the I
component and A(el) is the number of atoms. In the scenario where only one class
of polluters is at work, the value $d$ should be the same (within the errors) 
for all the involved elements, whatever pair of species we are considering. In
particular, this must be true for couples of elements involved in H-burning at
different temperatures, as Na-O, Mg-Al, and Al-Si.

We checked whether this happened for all the three components with intermediate
composition (P2, I1, and I2), deriving the quantity $d$ from all possible
combinations. The results are listed in Table~\ref{t:dil28}, where the dilution
fraction for different elements are listed..

\begin{figure}
\centering
\includegraphics[scale=0.40]{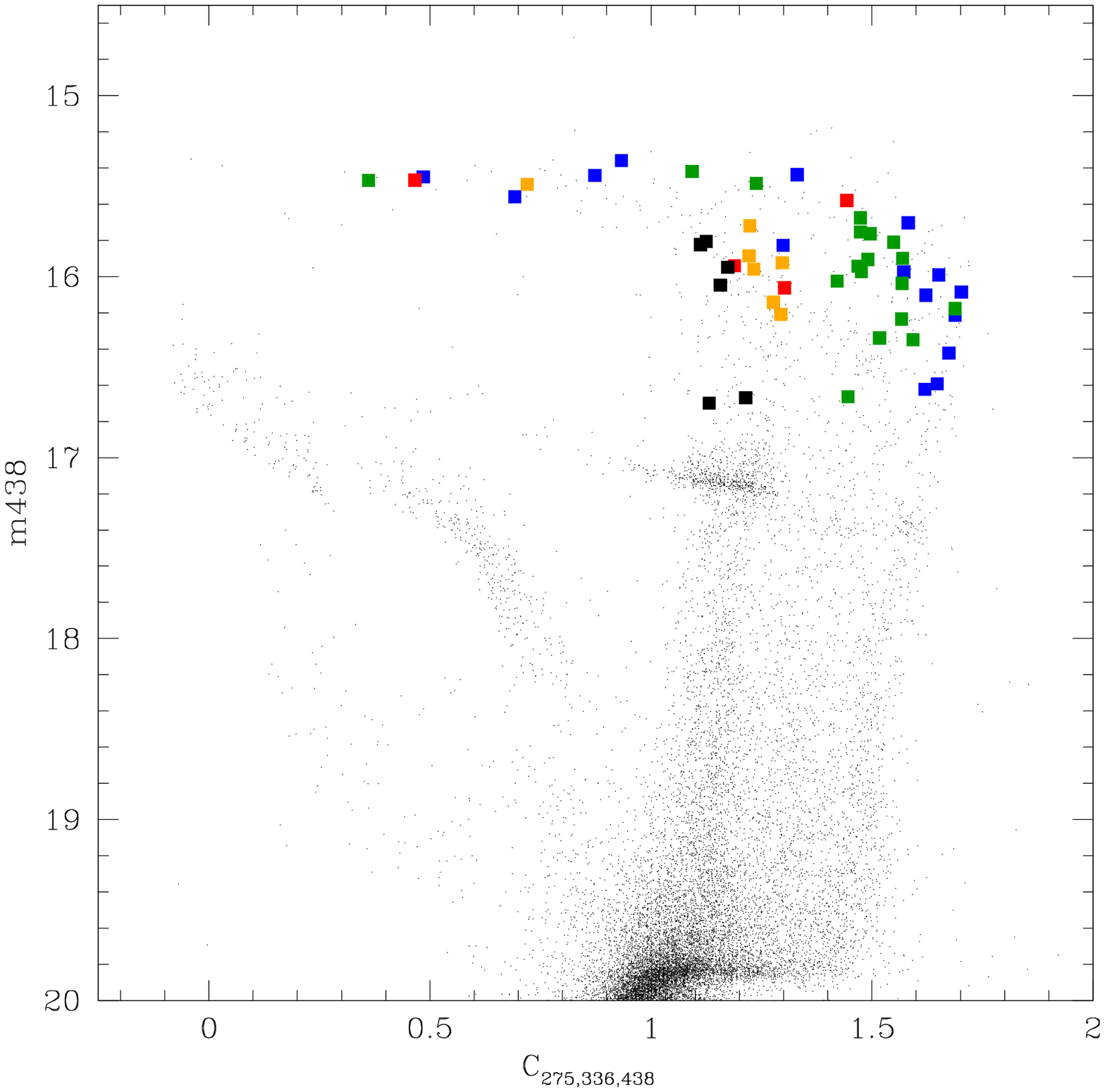}
\caption{Diagram using  
the pseudo-colour $C_{275,336,438}$ against $F438W$. The stars in our
spectroscopic sample are plotted using larger, filled symbols and adopting the
same colour scheme of previous figures (P1: blue, P2: green, I1: red, I2:
orange, and E: black).}
\label{f:photo}
\end{figure}

q\begin{figure}
\centering
\includegraphics[bb=284 146 584 715, clip, scale=0.52]{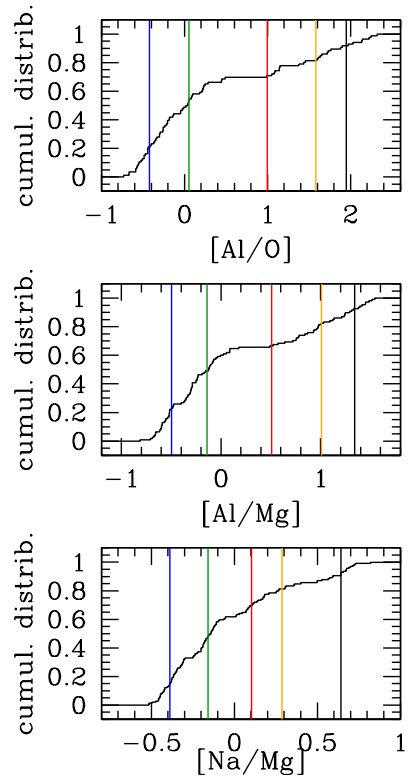}
\caption{Cumulative distribution of the [Al/O], [Al/Mg], and [Na/Mg] abundance 
ratios in NGC~2808 (from top to bottom). The vertical lines are traced at the
average values for the groups P1,P2,I1,I2,E using the same colour coding as in
previous figures (blue, green, red, orange, and black, respectively).}
\label{f:cumu3} \end{figure}

\setcounter{table}{4}
\begin{table*}
\centering
\caption[]{Dilution values for the components P2, I1, and I2}
\begin{tabular}{lcccc}
\hline
group & from Na-O & from Na-O & from Mg-Al  & from Mg-Al \\
\hline
      &     O           &     Na          &     Mg          &    Al           \\
P2    & $d=0.34\pm0.04$ & $d=0.18\pm0.02$ & $d=0.15\pm0.04$ & $d=0.04\pm0.01$ \\
I1    & $d=0.79\pm0.04$ & $d=0.45\pm0.04$ & $d=0.38\pm0.04$ & $d=0.36\pm0.07$ \\
I2    & $d=0.92\pm0.05$ & $d=0.54\pm0.08$ & $d=0.73\pm0.04$ & $d=0.64\pm0.07$ \\
\hline
group & from O-Mg & from O-Mg & from Na-Al  & from Na-Al \\
\hline
      &     O          &     Mg          &     Na          &    Al           \\
P2    & $d=0.34\pm0.04$ & $d=0.15\pm0.03$ & $d=0.18\pm0.02$ & $d=0.04\pm0.01$ \\
I1    & $d=0.79\pm0.04$ & $d=0.37\pm0.04$ & $d=0.48\pm0.04$ & $d=0.36\pm0.07$ \\
I2    & $d=0.92\pm0.05$ & $d=0.72\pm0.04$ & $d=0.53\pm0.06$ & $d=0.64\pm0.07$ \\
\hline
group & from Na-Mg & from Na-Mg & from O-Al  & from O-Al \\
\hline
      &     Na          &     Mg          &     O          &    Al           \\
P2    & $d=0.17\pm0.02$ & $d=0.16\pm0.03$ & $d=0.34\pm0.04$ & $d=0.05\pm0.01$ \\
I1    & $d=0.44\pm0.04$ & $d=0.37\pm0.04$ & $d=0.77\pm0.05$ & $d=0.38\pm0.08$ \\
I2    & $d=0.50\pm0.06$ & $d=0.74\pm0.04$ & $d=0.91\pm0.07$ & $d=0.54\pm0.07$ \\
\hline
group & from Si-O & from Si-O & from Na-Si  & from Na-Si \\
\hline
      &     O           &     Si          &     Na          &    Si           \\
P2    & $d=0.34\pm0.04$ &                 & $d=0.17\pm0.02$ &                 \\
I1    & $d=0.79\pm0.04$ & $d=0.26\pm0.08$ & $d=0.44\pm0.08$ & $d=0.27\pm0.07$ \\
I2    & $d=0.92\pm0.05$ & $d=0.43\pm0.17$ & $d=0.50\pm0.06$ & $d=0.49\pm0.11$ \\
\hline
group & from Si-Mg & from Si-Mg & from Si-Al  & from Si-Al \\
\hline
      &     Mg           &     Si          &     Al          &    Si           \\
P2    & $d=0.16\pm0.03$ &                 & $d=0.04\pm0.01$ &                 \\
I1    & $d=0.37\pm0.03$ & $d=0.27\pm0.04$ & $d=0.36\pm0.07$ & $d=0.25\pm0.07$ \\
I2    & $d=0.74\pm0.04$ & $d=0.49\pm0.11$ & $d=0.64\pm0.07$ & $d=0.52\pm0.10$ \\
\hline
\hline
\end{tabular}
\label{t:dil28}
\end{table*}

From this exercise we conclude that the fraction of material with E-like
composition decreases along the sequence I2--I1--P2. This decrease is
essentially what we are seeing
when considering the average abundances in these three components (e.g. in 
Fig.~\ref{f:a18}). Moreover, by comparing the values for the individual 
species, we found that the $d$ values are different, beyond the associated
errors. As for the much more simple case of NGC~6752, where at least two classes of
polluters were required, our findings show that the composition of the P2, I1,
and I2 groups in NGC~2808 must be produced by different polluters. A simple
dilution model with only a class of polluters does not seem to be enough to
reproduce the observed pattern of the intermediate populations in NGC~2808. 

Finally, we tried to link the information coming from our spectroscopic sample
to what is derived on photometric grounds. In particular,  we used the 
preliminary release of the HST UV Legacy program GO-13297 (see Piotto et al.
2015), downloading the data for NGC~2808 from
http://groups.dfa.unipd.it/ESPG/treasury.php. In fact, the filter combination
({\em F275W, F336W}, and $F438W$) used in that program is particularly efficient
in detecting variations in C, N, and O (see Piotto et al. 2015). Milone et al.
(2015) employed this dataset, in combination with previous HST data, to study
NGC~2808 and deduced the presence of five populations. The correlation between
the photometrically and spectroscopically defined multiple population was
discussed in Milone et al. (2015) and Carretta (2015), but we add here more
data.

Fig.~\ref{f:photo} shows the diagram obtained plotting the pseudo-colour
$C_{275,336,438}=F275W-2 \times F336W + F438W$ against $F438W$; the figure is
very similar to what is shown in Marino et al. (2017, their Fig.~7) even if we
did not apply any cut on photometric quality. We cross-matched the HST UV
catalogue with ours and found 62 stars in common (50 of which have all the three
magnitudes). They are plotted in Fig.~\ref{f:photo} using different colours,
according to their spectroscopically defined population. 
Neglecting the brightest portion of the RGB, where this pseudo-colour is no more
efficient in separating the populations, the stars cluster along the separate 
RGBs, with distinct populations falling on different RGBs.

\section{Cluster analysis}

More information can be inferred from the cumulative distributions of abundance
ratios along the main anti-correlations among proton-capture elements, shown in
Fig.~\ref{f:cumu3}.

Starting with the [Na/Mg] ratio (used by Carretta 2015 to classify the five
groups), we plotted vertical lines at the position given by average values for
the populations P1, P2, I1, I2, and E, using the same colour coding as before.
The five average values bracket four regions where the cumulative distribution
is flat, i.e. the number of stars there does not increase. In other words, the
mean values corresponding to the different sub-populations in NGC~2808 
bracket $gaps$ in the observed Na-Mg anticorrelation. The same also holds for
the ratios [Al/Mg] and [O/Na].

\begin{figure*}
\centering
\includegraphics[scale=0.60]{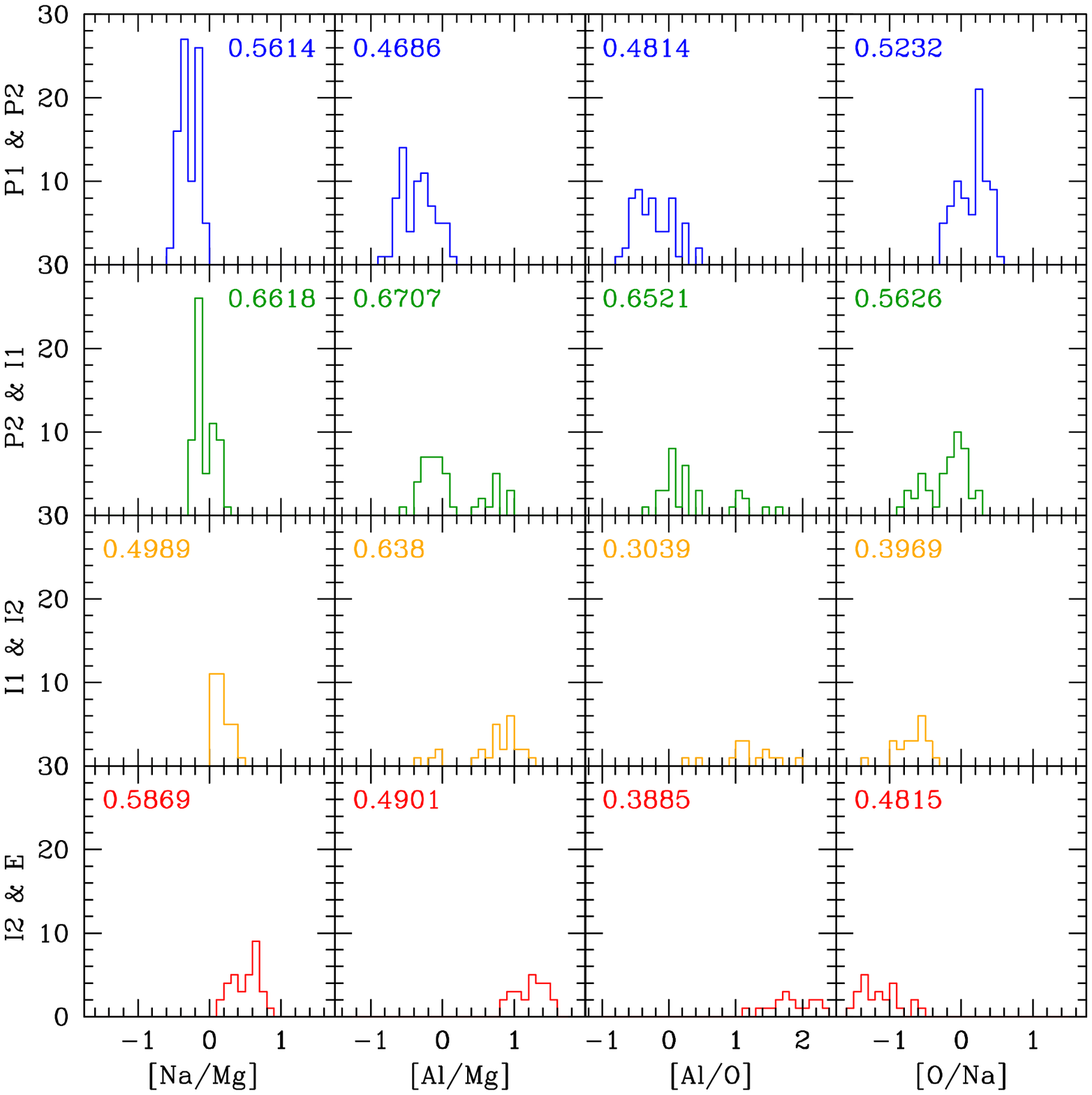}
\caption{Observed distributions along the Na-Mg, Al-Mg, Al-O, and O/Na
anticorrelations for the sub-samples P1+P2, 
P2+I1, I1+I2, and I2+E in NGC~2808. In each panel the value of the bimodality
coefficient BC is labelled.}
\label{f:plot16}
\end{figure*}

To quantify the reality of the gaps, we
considered for each combined pair of sub-samples the Sarle's bimodality 
coefficient (BC), to be computed from the skewness and the excess kurtosis, both 
corrected for sample bias (SAS Institute Inc.1990). The distributions are 
shown in the panels of Fig.~\ref{f:plot16}. In each panel we also list the
value of the BC.

When compared to the critical value $5/9 \sim 0.555$ expected for a uniform 
distribution, higher values point toward bimodality and lower values toward
unimodality. Whereas using this test on the whole sample of stars available in
NGC~2808 always indicates that all distributions are not unimodal, the results
are more uncertain when splitting the total sample in sub-groups that should be
bracketing the gaps defined by [Na/Mg] in Carretta (2015).
For example, this coefficient formally does not allow to recognize the
distribution P1-P2 in [Al/Mg], [O/Na], and [Al/O] as bimodal, even if the
bimodality is clearly evident by eye for at least the first two (first row in
Fig.~\ref{f:plot16}, second and fourth panel). Hence, maybe this is not the best
test to quantify the distribution of stars along the anti-correlations.

Next, we resort to cluster analysis using the $k-$means algorithm
(Steinhaus 1956; MacQueen 1967) as implemented in the $R$ statistical package (R
Development Core Team 2011; http://www.R-project.org), as done in Carretta et
al. (2012a,c). 
We began by selecting the same ratios [Na/Fe] and [Mg/Fe] used by Carretta
(2015) to define the five discrete populations on the RGB in NGC~2808.

\begin{figure}
\centering
\includegraphics[scale=0.40]{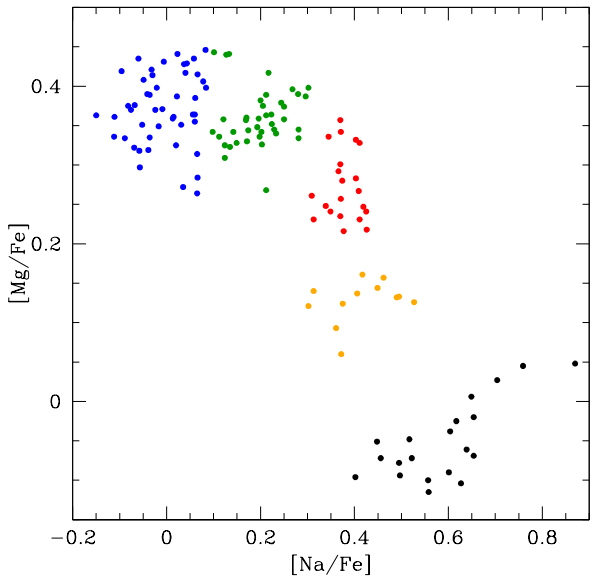}
\caption{Results of the cluster analysis using the $k-$means algorithm and
the selected ratios [Na/Mg], [Mg/Fe].}
\label{f:provakmeans}
\end{figure}

The algorithm retrieves five groups (Fig.~\ref{f:provakmeans}, which is an 
objective quantification of the populations with distinct chemistry detected by
Carretta (2015). The only difference is that four stars of the P2 group are now
classified as P1 stars and three stars that were assigned by eye as P1 now move
to the P2 group (compare this figure to Figure 8 in Carretta 2015).  The other populations are identical to
those defined by eye in Carretta (2015). This is likely because the algorithm
favours a subdivision based especially on Na, whose variation is much larger
than that of Mg.

However, for NGC~2808 we now have $six$ proton-capture elements available, hence
we tried a more robust cluster analysis using all stars with CN, O, Na, Mg, Al,
and Si abundances. By using all the involved species, we found that only three
main groups have a robust justification: they correspond essentially to the P,
I, and E components as defined in Carretta et al. (2009a). However, the
intermediate (I) and extreme (E) groups are not homogeneous, in abundance.

\begin{figure}
\centering
\includegraphics[scale=0.40]{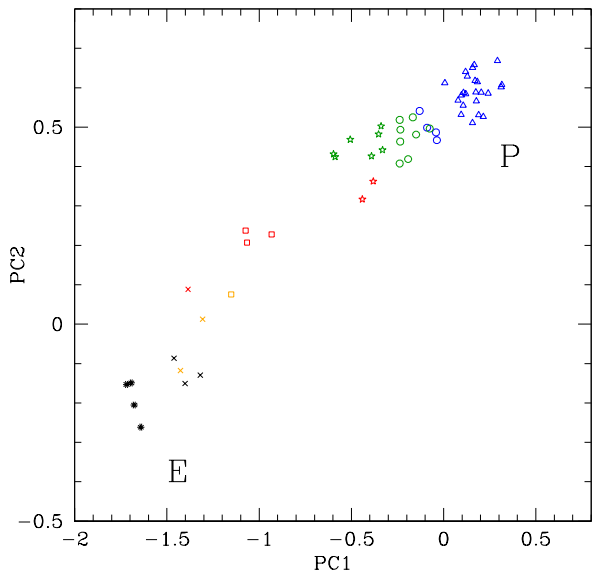}
\caption{Results of the cluster analysis using the $k-$means algorithm 
only on stars with the full set of CN, O, Na, Mg, Al, and Si abundances. Colour
coding is as in previous figures, whereas different symbols indicate the groups
found by cluster analysis.}
\label{f:sara}
\end{figure}

In order to obtain homogeneous populations the best sub-division is into $six$
groups. Results are displayed in Fig.~\ref{f:sara}, showing the relation
between the six groups individuated by the $k-$means algorithm and the five
components discussed so far. The colour coding is
the same as in previous figures and refers to the five populations determined by
Carretta (2015). Different symbols indicate the six components found by the
cluster analysis. In general the correspondence is good, but there
are exceptions. The homogeneous groups are essentially separated according to
the PC1 component that alone account for $\sim 82\%$ of the variance.
One possible way to look at these results is that rather than a the linear
combination of two vectors, we are looking at the non-linear  combination of a
single vector (a combination of PC1 and PC2),  that is the effect of one single
physical process.  Therefore, while numerically, by definition, PC1 and PC2 are
orthogonal, that is actually not the case from a physical point of view.

\setcounter{table}{5}
\begin{table*}
\centering
\caption[]{Average abundances of CN, Al and other proton-capture elements in 
the five groups in NGC~2808}
\begin{tabular}{lrccrcc}
\hline
group & n  & [Al/Fe] &    rms  &  n &  [CN/Fe]  &  rms     \\
\hline
P1    & 39 & $-$0.114$\pm 0.023$&0.141&33   &$-$0.283$\pm 0.014$ &0.082\\
P2    & 28 &   +0.206$\pm 0.032$&0.168&22   &$-$0.221$\pm 0.017$ &0.078\\
I1    & 16 &   +0.782$\pm 0.097$&0.387&12   &$-$0.136$\pm 0.028$ &0.098\\
I2    & 10 &   +1.136$\pm 0.030$&0.095& 8   &$-$0.114$\pm 0.035$ &0.100\\
E     & 15 &   +1.292$\pm 0.029$&0.113&14   &$-$0.017$\pm 0.017$ &0.101\\
\hline
\hline
group & n  & [O/Fe] &    rms  &  n &  [Na/Fe]  &  rms     \\
\hline
P1    & 42 &   +0.308$\pm 0.009$&0.058&46   &$-$0.005$\pm 0.010$ &0.067\\
P2    & 35 &   +0.154$\pm 0.015$&0.090&40   &  +0.188$\pm 0.011$ &0.068\\
I1    & 15 & $-$0.216$\pm 0.032$&0.124&21   &  +0.378$\pm 0.007$ &0.034\\
I2    &  7 & $-$0.447$\pm 0.082$&0.217&12   &  +0.414$\pm 0.021$ &0.072\\
E     & 17 & $-$0.656$\pm 0.039$&0.161&20   &  +0.592$\pm 0.025$ &0.112\\
\hline
group & n  & [Mg/Fe] &    rms  &  n &  [Si/Fe]  &  rms     \\
\hline
P1    & 46 &   +0.384$\pm 0.006$&0.041&46   &  +0.265$\pm 0.004$ &0.026\\
P2    & 40 &   +0.346$\pm 0.006$&0.035&40   &  +0.262$\pm 0.004$ &0.026\\
I1    & 21 &   +0.274$\pm 0.010$&0.044&21   &  +0.309$\pm 0.006$ &0.026\\
I2    & 12 &   +0.127$\pm 0.008$&0.028&12   &  +0.346$\pm 0.011$ &0.038\\
E     & 20 & $-$0.050$\pm 0.011$&0.050&20   &  +0.390$\pm 0.008$ &0.036\\
\hline
\hline
\end{tabular}
\label{t:medieNAl}
\end{table*}

\section{Discussion and conclusions}

In the present work we derived homogeneous Al abundances for 108 RGB stars in
NGC~2808, as well as estimates of the CN content in 89 giants. These elements
join the large dataset of proton-capture elements obtained from spectra
in this massive globular cluster, and both species nicely fit in the
overall pattern of (anti)correlations produced by the network of proton-capture
reactions in H-burning at high temperature (Fig.~\ref{f:light3t28def} and
Fig.~\ref{f:light4t28def}).
We found that Al abundances span an interval of almost 2 dex, and are 
anti-correlated to the Mg abundances that, in turn, are depleted by almost 
0.6 dex with respect to the typical overabundances present in metal-poor halo
stars in most extreme cases. 

The improved statistics allow us to conclude that a unique class of polluters is
not enough to reproduce the entire set of anti-correlations among
proton-capture elements CN, O, Na, Mg, Al, and Si. When ejecta from the most massive
FG stars are mixed up with variable amount of pristine matter in a dilution
model our data show that different polluters contribute to the formation of the
discrete populations P2, I1, and I2, showing intermediate composition.
This is in agreement with the observational constraints derived from Li
abundances in lower RGB stars in NGC~2808 by D'Orazi et al. (2015). They found
that all the Al-poor, FG stars in their sample (this would correspond to groups
P1 and P2 in our nomenclature) are also Li-rich. Among the SG Al-rich stars, 
beside the extreme population devoid of Li, as expected from the high
temperatures producing Al enhancement in the FG polluters, they also detected
a sample of stars having Li abundances similar to those of
FG stars. About two-thirds of Al-rich stars are also Li-rich and about one third
Li-poor. Since the lower and upper RGB stars are drawn from the same population, 
we should also expect that two-thirds of the combined samples of I1+I2+E stars
are Li-rich. These observations suggests that at least two different kinds
of polluters were active in NGC~2808, one of them being able to produce some
amount of fresh lithium. 

A one-to-one comparison between the abundances of Li and of the elements
involved in the p-capture processes is hampered by the lack of stars in common between
the present sample and the one analyzed by D'Orazi et al. However, they can be
compared in some way using as a common ground the information from $HST$
photometry by Milone et al. (2015). 

\begin{figure}
\centering
\includegraphics[scale=0.40]{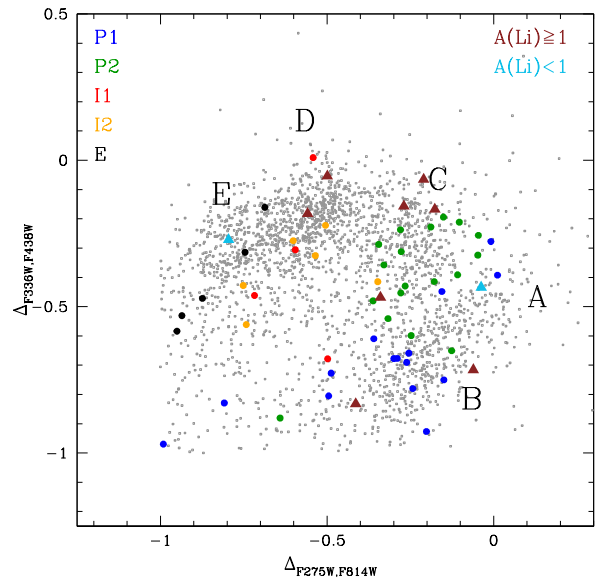}
\caption{CNO-two-colour diagram (D'Antona et al. 2016, Milone et al. 2015: grey
points). Larger filled circles are stars with our spectroscopic abundance determination,
colour-coded according to the groups in Carretta (2015). Large black letters
indicate the five photometric groups. Brown and cyan filled triangles are
Li-rich and Li-poor stars from D'Orazi et al. (2015).}
\label{f:cromoLi}
\end{figure}

Using stars in common between our sample and the preliminary release of $HST$
photometry available to us (see Sect. 4) we reproduced the CNO two-colours
diagram in Fig.~\ref{f:cromoLi}, where we superimposed our stars, divided into the 5 chemical groups 
and the classification from photometry. From this figure and the discussion in
Carretta (2015), it is straightforward enough to identify our P1 group with the
group B of Milone et al. (2015). This group should have a normal He content (see 
also D'Antona et al. 2016). The P2
component coincides with the photometric group C, and the populations I1 and E
can be identified with group D and E (Carretta 2015, Milone et al. 2015,
D'Antona et al. 2016). The spectroscopic group I2 is not easily associated to a
specific photometric group, but the spreads in the photometry - exceeding the
analysis errors - may even hide more sub-populations.

We found that 10 stars from D'Orazi et al. (2015) also have all the required 
magnitudes to be plotted in Fig.~\ref{f:cromoLi}. The 8 Li-rich stars are
distributed in the P1 group (2 stars, group B with Y=0.278 in Milone et al.,
the P2 group (4 stars, group C with Y=0.280), and in the region populated by I1
and I2 groups (2 stars, group D with Y=0.318). 
All estimates of the He mass fraction given in this section are
from Milone et al. (2015).
Taking into account the very
limited sample, this correspond to a fraction $80\pm28\%$, consistent with the
above estimate of Li-rich stars. Two stars in this sample have moderately low
Li abundances. One of them lie within the locus of the
photometric group A (Y=0.243), which has no counterpart in the spectroscopic
sample, whereas the other star is in the E component (both spectroscopic and
photometric groups, with Y=0.367).

Summarizing, our new data confirm that stars along the RGB in NGC~2808 are
segregated into the five distinct groups defined in Carretta (2015) using
only the Mg, Na abundances. This is particularly evident in the Mg-Al
anti-correlation. The Si-Al correlation
(Fig.~\ref{f:light3t28def}, lower right panel) does mirror the Si-Mg
anti-correlation (see Carretta 2015, his fig. 11) and confirms the existence in
this cluster of the leakage from Mg to Si (Karakas and Lattanzio 2003) through a
reaction only efficient at temperatures exceeding 65 MK (Arnould et
al. 1999).

Together, the high values of Al and the large depletion in Mg reached in the
component with the most extreme composition put strong constraints on the
nature of possible polluters providing nuclearly processed matter. As discussed
in Carretta (2014), in the FRMS scenario there is only a narrow time interval
when such high inner temperatures are reached, just at the end of the
main-sequence evolution of massive, rotating stars. However, in this phase also
Na is destroyed, whereas this species reaches its highest value in the E
component. D'Antona et al. (2016) go further and state that the the observed Mg
depletion completely rules out any scenario based on massive stars, either
single or in interacting binaries (de Mink et al. 2009).

On the other hand, even the AGB scenario seems to face several issues when 
compared to the observational data. Carretta (2014) already pointed out a 
possible inconsistency: the candidate polluters providing the maximum depletion
in Mg are the AGB stars with  5-6 $M_\odot$ initial masses (D'Ercole et al.
2012), whose He content is however below the threshold required for the onset of
a deep-mixing able to push oxygen abundances down to [O/Fe]$\simeq -1$ dex, as
instead observed in very Mg-poor stars in NGC~2808.

\begin{figure}
\centering
\includegraphics[scale=0.40]{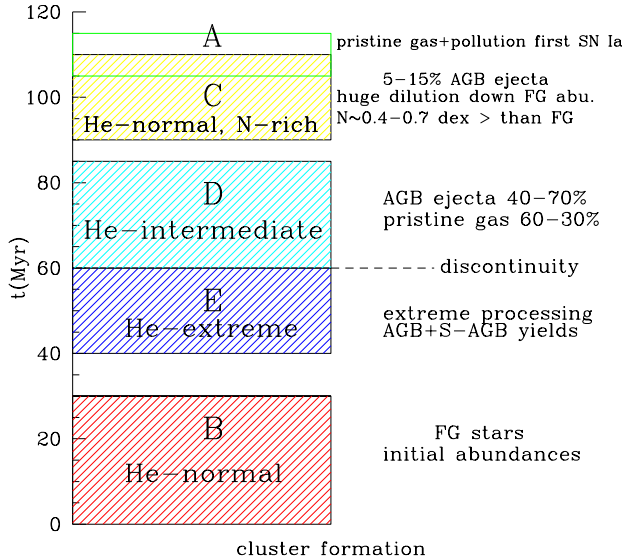}
\caption{Sketch of the temporal sequence proposed by D'Antona et al. (2016) to
explain the discrete populations in NGC~2808. The dashed line indicate a
chemical discontinuity between the photometric groups E and D owing to a sudden
dilution with pristine matter. The ordinate is the time (in million years) since
the first burst of star formation in the cluster.}
\label{f:sequenzaBEDCA}
\end{figure}

How these new data can help to understand the formation scenario in
this cluster?
Recently, D'Antona et al. (2016) proposed a temporal sequence to explain
the complex pattern of multiple, and discrete, populations in NGC~2808. We
sketched their scenario in Fig.~\ref{f:sequenzaBEDCA}.

In their scenario the 
groups B and C share the same normal He content, but only the first is actually
composed by first generation stars, with typical abundances of halo
stars. The C component is seen as one of the last formed in the cluster, some
90-110 Myr after the first burst of star formation. As a consequence, the action
of lowest mass AGB stars, able to show the effect of third dredge-up episodes,
is visible in this group through an enhanced abundance of N, higher by 0.4-0.7
dex than the N content of group B. This large difference in N provides the
different location of groups B and C in the CNO-two-colours diagram (see D'Antona
et al. 2016) that we reproduced in Fig.~\ref{f:cromoLi}.

However, there seems to be no correspondence for this difference in our
spectroscopic data for groups P1 and P2. If we read from the generalised
histogram of $\delta S_{CN}$ in NGC~2808 (figure 6 in Norris and Smith 1983) 
the values corresponding to the most CN-poor populations and we use the
calibration in Fig.~\ref{f:dscn} to convert the values to [CN/Fe] ratios onto our
present scale, we found a difference of only 0.13 dex between components P1 and
P2, leaving aside an arbitrary zero point due to the unknown C
abundances. Moreover, from  Table~\ref{t:medieNAl}, on average the two groups
differ by 0.06 dex only in their CN content. 

From spectroscopy, apparently we do not find a large difference in CN between 
the populations with the most primordial composition. This finding is quite
puzzling, since the photometric He-normal, N-rich group is numerically
significant, with an estimated fraction $26.4 \pm 1.2\%$ (Milone et al. 2015).

Turning to the search for a general scenario, the present results imply that in
NGC~2808 different classes of polluters were likely at work,  producing a
syncopated, discrete distribution of different populations (at least five, but
probably more) along the main (anti)correlations among the proton-capture
elements.

Our observations, therefore, tend to exclude massive stars as responsible for
most of the intermediate populations, in
NGC~2808, since these stars may only act at the very beginning of the cluster
evolution, in the first few million years (and moreover they cannot produce
the lithium observed in SG stars by D'Orazi et al. 2015). After the massive
SN II went off, clearing the gas reservoir, the only way to produce more than
two discrete populations in this scenario would be to periodically inject new
and variable amounts of pristine, unpolluted gas to be mixed. However, the gas
with primordial composition would find no more polluted gas after the SN II
epoch. It seems difficult to justify more than two discrete populations within
the FRMS framework. However, our data cannot exclude that the FRMS were
responsible for the first polluted SG in this cluster.

In presence of more than two, and discrete, stellar populations in GCs, the AGB
scenario has the advantage of providing a range of possibly contributing stellar
masses. In turn this offers different processed yields released at (even very)
different epochs and with different temperatures involved. Although
inconsistencies related to the nucleosynthesis still exist, as outlined above, 
this scenario provides a natural way to explain SG stars where we see the outcome
of the complete Mg-Al-Si, a SG produced by less massive polluters where the
Mg-Al cycle is active, but without touching on Si, down to a SG where we only
see O depletion and Na enhancement, generated from still less massive stars. 
Moreover, in this framework it is relatively easy to account for the observational
constraints derived from lithium, since not all the AGB stars are good producers
of Li.

D'Antona et al. (2016) exploited these properties by combining a
temporal sequence of star formation episodes (sketched in
Fig.~\ref{f:sequenzaBEDCA}) with dilution events. By tuning star formation
bursts and dilution, they showed how the chemical composition of the five
groups individuated in NGC~2808 may be reproduced. However, the conflicting
evidence concerning the N abundances as derived from spectroscopy and
photometry possibly uncovered in our present work
still remains an open issue.

Recently, Bekki et al. (2017) claimed that in the scenario by D'Antona et
al. no physical mechanism was clearly introduced to stop and re-start several
time the star formation and gas accretion. Bekki et al. proposeed a model 
to interrupt the different bursts of star formation and were able to
reproduce up to 5-6 discrete populations using a time-evolving Initial Mass
Function. On the other hand, their model does not predict chemical
evolution independently, but simply adopts AGB yields from Ventura et al. (2009,
2011). 

While it is desirable that more and more physically self-consistent
chemo-dynamical models are produced, it is also clear that the peculiar globular
cluster NGC~2808 is one of the best (and toughest) benchmark to
test any scenario for the origin and the evolution of multiple populations in
GCs.

\begin{acknowledgements}
This research has made use of the SIMBAD database (in particular  Vizier),
operated at CDS, Strasbourg, France, of the NASA's Astrophysical Data System, 
and {\sc TOPCAT} (http://www.starlink.ac.uk/topcat/). This work made use of 
R: A language and environment for statistical computing. R Foundation for
Statistical Computing, Vienna, Austria. ISBN 3-900051-07-0, URL
http://www.R-project.org.
\end{acknowledgements}

\end{document}